# Fitted avatars: automatic skeleton adjustment for self-avatars in virtual reality

Jose Luis Ponton[1] · Víctor Ceballos[2] · Lesly Acosta[3] · Alejandro Ríos[1] · Eva Monclús[1] · Nuria Pelechano[1]

**Abstract**
In the era of the metaverse, self-avatars are gaining popularity, as they can enhance presence and provide embodiment when a user is immersed in Virtual Reality. They are also very important in collaborative Virtual Reality to improve communication through gestures. Whether we are using a complex motion capture solution or a few trackers with inverse kinematics (IK), it is essential to have a good match in size between the avatar and the user, as otherwise mismatches in self-avatar posture could be noticeable for the user. To achieve such a correct match in dimensions, a manual process is often required, with the need for a second person to take measurements of body limbs and introduce them into the system. This process can be time-consuming, and prone to errors. In this paper, we propose an automatic measuring method that simply requires the user to do a small set of exercises while wearing a Head-Mounted Display (HMD), two hand controllers, and three trackers. Our work provides an affordable and quick method to automatically extract user measurements and adjust the virtual humanoid skeleton to the exact dimensions. Our results show that our method can reduce the misalignment produced by the IK system when compared to other solutions that simply apply a uniform scaling to an avatar based on the height of the HMD, and make assumptions about the locations of joints with respect to the trackers.

✉ Nuria Pelechano
npelechano@cs.upc.edu

Jose Luis Ponton
jose.luis.ponton@upc.edu

Víctor Ceballos
victor.ceballosinza@kaust.edu.sa

Lesly Acosta
lesly.acosta@upc.edu

Alejandro Ríos
arios@cs.upc.edu

Eva Monclús
emonclus@cs.upc.edu

1    Research Center for Visualization, Virtual Reality
     and Graphics Interaction - ViRVIG, Universitat Politècnica
     de Catalunya, Barcelona, Spain

2    Visual Computing Center - VCC, King Abdullah University
     of Science and Technology, Thuwal, Saudi Arabia

3    Estadística i Investigació Operativa - EIO, Universitat
     Politècnica de Catalunya, Barcelona, Spain

## 1 Introduction

Head-Mounted Displays (HMD) offer an immersive experience that can greatly enhance presence in a virtual environment. The main problem when experiencing a virtual environment through this technology is that users cannot see their own bodies when they look down or try to interact with the environment. This creates the strange feeling of being a *floating head* without a body, which, in addition to being disturbing, can introduce other undesirable effects, such as increased motion sickness, or lack of correct perception of distances and sizes.

For a long time, this issue has been mostly ignored in applications such as video games, architecture visualization, education, and training. One of the reasons is that having a realistic-looking body following the user's movement requires very costly equipment, such as full-body scanners and high-end motion capture systems.

As HMDs are becoming more affordable and more users have access to this technology, it is necessary to explore solutions to introduce low-cost self-avatars into any virtual environment. Currently, there are many tools that can be used to rapidly create avatars such as MakeHuman[1] and CharacterGenerator.[2] They produce humanoid meshes, each with a biped skeleton and a rig, ready to be animated in any engine.

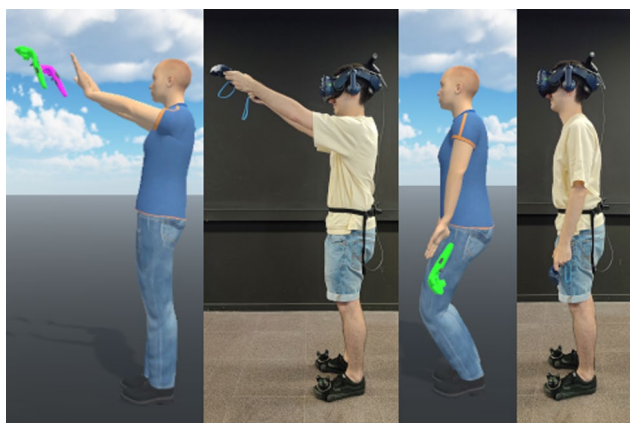

**Fig. 1** Problems when the dimensions of the user's limbs differ from those of the avatar. On the left, controllers appear to fly away. On the right, the avatar's knees are bent when the user stands up

In order to have an avatar that accurately follows the user's movements, it is necessary to track the user and apply real-time inverse kinematics (IK) to animate the avatar in the virtual environment. There are current solutions for real-time IK that can be used in a game engine. The problem, however, is that an IK solver will not provide accurate movement unless the avatar proportions correspond to those of the user. In case of a dimension mismatch, there will be visible discrepancies such as the avatar's knees being bent while the user is standing upright, or the controllers appearing to fly away from the avatar's hand (see Fig. 1). This lack of adjustment is especially problematic when having a self-avatar because humans are very aware of their own body position; any mismatch with the avatar can be easily noticeable. For this reason, most high-quality mocap systems, such as Xsens, and several IK solutions for game engines, require the user to manually measure the length of their limbs and enter the information into the system in order to adjust the self-avatar skeleton correctly.

In this paper, we present a method to fit an avatar to the dimensions of a user rapidly. We use the HTC VIVE Head-Mounted Display (HMD), with two VIVE wand controllers (one for each hand), and three additional VIVE trackers, attached with straps to the user's waist and feet. The user needs to follow a simple sequence of movements to capture the exact location of relevant joints and lengths of limbs. Once the avatar has been adjusted, we can animate it in real time using only those six trackers. The main advantage of this system is that the user can easily and quickly set up the avatar adjustment without any external assistance, using low-cost technology.

The contribution of this paper consists of a low-cost and easy-to-setup system that could be incorporated into any VR application where self-avatars are needed, without needing external assistance. Our results show that our method can quantitatively reduce the misalignment produced by the IK solution when compared to an avatar with a uniform scaling based on the user height. Participants experience higher embodiment and were less aware of pose artifacts when using our fitted avatar instead of a uniform scaling, but only when the initial avatar proportions differed from the user's proportions.

## 2 Related work

### 2.1 Motion capture

High-end solutions for mocap include camera-based systems such as Vicon, or inertial-based systems such as Xsens. There are also lower-cost solutions based on RGB cameras (Mehta et al. [2020]; Shi et al. [2020]), or depth cameras (Wei and Zhang [2012]). However, the use of external cameras requires the user to stay within the field of view of the cameras, which can be challenging when the user is immersed in a Head-Mounted Display. Most HMDs nowadays provide at least two controllers to interact with the virtual world. These devices provide the position and orientation of the head and hands of the user, which can be used to animate the upper body of an avatar. In the case of HTC VIVE, Zeng et al. [2022] include a few additional trackers, which, combined with an IK solver, can provide full self-avatar movement. They uniformly adjust the avatar's size in accordance with the user's height, and thus assume that the avatar always has the same ratios between bone lengths as the user. Therefore, their solution does not consider having a variety of body proportions as in our work. We refer to the survey by Aristidou et al. [2018] for an extensive survey on IK solutions.

There are currently several options to carry out IK using a few HTC VIVE trackers. Unity's built-in IK solves each limb individually using four end-effectors (feet and hands), so two controllers and two HTC trackers can be used to animate the self-avatar, while the HMD orientation is applied to the avatar head. Unfortunately, this solution is very limited, as it simplifies the IK solution to each limb, and will not adjust the back of the avatar to lean toward the tracker if it becomes out of reach. Also, since there is no end-effector for the avatar's head, if the person bends forward, the avatar will simply rotate its head accordingly, but will not modify its back joins. FinalIK[3] computes joint rotations for all body parts to minimize the distance between the end-effectors and the HTC trackers, and includes an IK handler for the head which provides torso rotations and bending. However, it is not easy to assign trackers to end-effectors, and the fixed offsets applied between them result in incorrect alignments with the user, thus leading to large errors in body rotations and the overall positioning of the skeleton.

## 2.2 Character generation

There are currently many tools to generate avatars, such as MakeHuman or Autodesk Character Generator. These tools allow the user to tune different parameters to generate avatars of different heights, sizes, and proportions. However, once an avatar has been built and imported to a game engine, it cannot be changed any further, and thus it is not possible to adjust its dimensions to those of the real user.

High-quality characters that perfectly match a user's size and appearance, can be created with expensive 3D scanners (Thaler et al. 2018), which are not accessible to the general public. And even if the external dimensions of the avatar are a perfect match, it is still impossible to have good correspondence with the inner skeleton, because this information cannot be captured with such technology.

The Virtual Caliper (Pujades et al. 2019) presented an easy-to-use framework to rapidly generate avatars that fit the shape of the user. They use the HTC VIVE wand controllers to take measurements of the user, and then they verify that they are linearly related to 3D body shapes from the SMPL body model (Loper et al. 2015). But even if the size of the avatar is correct in terms of the width of arms, legs, torso, etc., it does not guarantee that the positions of the avatar's joints match those of the user, and this has a direct impact on achieving realistic positioning of limbs during motion. The most noticeable one is typically the shoulder location since it affects both the distance that our hand can reach, and how the arm bends.

Regardless of the technology used to generate the avatars, the problem of finding a good adjustment between the trackers and the avatar skeleton remains. A wrong mismatch may not be crucial when avatars are being used as characters that we interact with in Virtual Reality (VR), but in the case of self-avatars, even the smallest misalignment turns into an incorrect body position for the avatar that our proprioceptive system can detect.

## 2.3 Embodiment

Having a self-avatar in VR can induce a sense of embodiment that makes the user feel as if the virtual avatar was their own body. Embodiment requires correct tracking of the participant to animate the virtual avatar (Slater et al. 1995), and synchronized movements (Gonzalez-Francoand Lanier 2017).

Kilteni et al. (2012) described the *sense of embodiment (SoE)* as the extent a participant feels the virtual body is their own. They propose to divide the *SoE* into three subcomponents that can be measured through questionnaires: *sense of self-location* (being inside a virtual body), *sense of agency* (being in control of the virtual body movements), and *sense of ownership* (appearance of the virtual body representing the user).

Several aspects need to be taken into account to achieve embodiment, such as appearance, visuotactile stimuli, point of view, and virtual body movements. Maselli and Slater (2013) showed that first-person perspective over a virtual body is essential for evoking the illusion of body ownership, even without visuotactile or sensorimotor cues.

Embodiment has been proven to be essential to enhance user experience (UX) and has implications for the way the user perceives the virtual environment and experiences presence. For example, Mohler et al. (2010) showed that having a self-avatar can improve distance judgment when observing a virtual environment with an HMD. This is crucial for VR applications in architecture, where the user needs to have a reliable representation of the space. In a later experiment, which portrayed Lenin giving a speech to Red Army recruits in Moscow, Slater et al. (2018) showed that presence was greatest when a participant was embodied as Lenin or as part of the audience rather than observing the same scene from a third-person point of view. Unlike Freud's experiment by Osimo et al. (2015), in which the whole body was tracked, when embodied as Lenin, only the head was tracked in real time. The movements of the virtual hands were prerecorded and participants noticed a lack of synchronization. Though they got a lower result in ownership, the *SoE* scores were still high. An interesting observation from this study is that when the participants noted the lack of synchrony with the hands, they attempted to match their movements to the Lenin virtual body.

Gonzalez-Franco et al. (2020) also examined the participants' behavior when observing asynchronous visual feedback. If there is a drift between the self-avatar and the physical body, the participants tend to fill the gap and reduce the offset. The authors call this the *self-avatar follower effect* and emerge from the brain's need to solve sensory conflicts. The higher the sense of ownership, the higher the need to reduce the offset.

Recently there has been an increasing interest in using collaborative virtual environments as a working platform for different fields. In these types of applications, two or more participants share a virtual environment to perform a certain task (Frost and Warren 2000; Andujar et al. 2018). These applications can greatly benefit from having virtual avatars to enhance nonverbal communication.

---

[3] RootMotion's FinalIK: http://root-motion.com/.

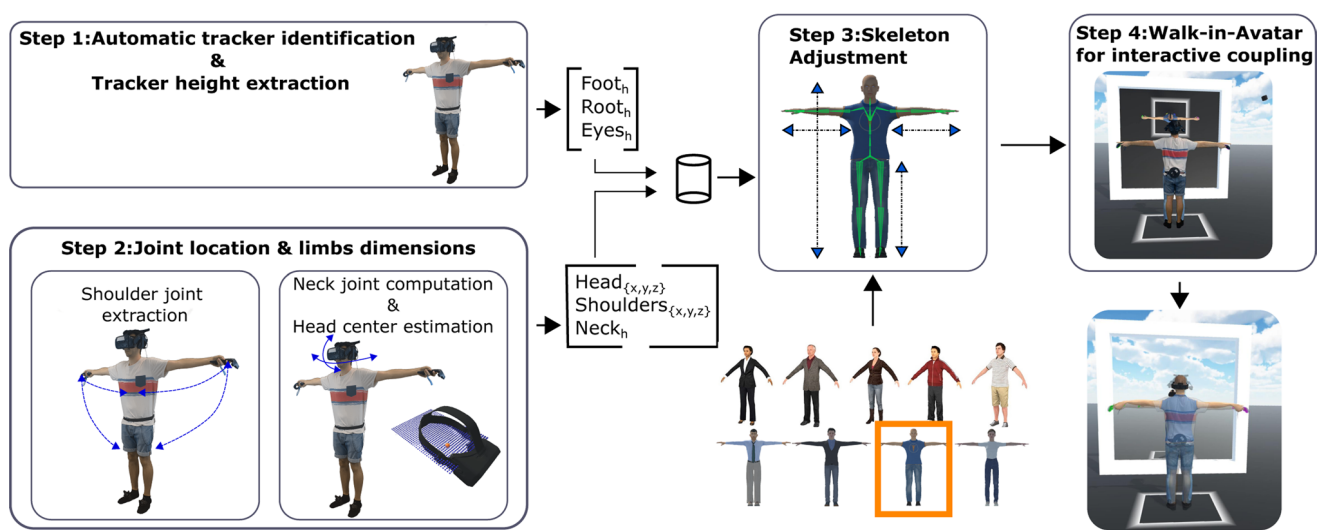

**Fig. 2** Framework setup. After choosing an avatar, the user stands on a T-pose to automatically identify the trackers and extract some measurements (Step 1). Next, the user follows a short sequence of exercises to automatically extract the location and dimensions of the user's joints and limbs (Step 2). Then the avatar skeleton is resized to match the user dimensions (Step 3). Finally, the user has to place himself inside the visualized avatar in order to compute the exact offsets between the avatar joints and the user trackers (Step 4)

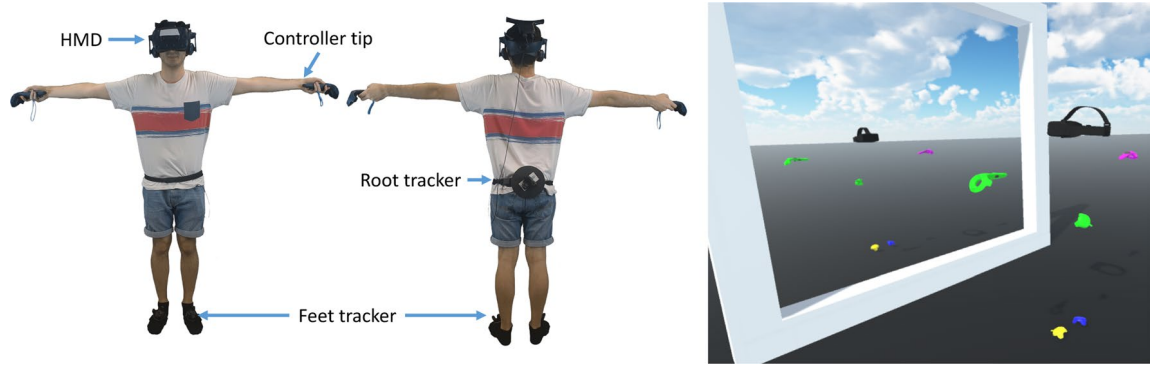

**Fig. 3** User set up with all the trackers, and the visualization when the user is in VR in front of the mirror

Ponton et al. (2022) presented a framework to compute exact offsets between six trackers located on the user body and the joints of its avatar. They observed that computing exact offsets, as opposed to default fixed ones, would improve the self-avatar movements leading to higher *SoE*. However, their avatars were simply adjusted in height to the user by applying a uniform scaling, which could not solve problems regarding the correct positioning of end-effectors (for example the hands) when animating the self-avatars.

When having a small number of trackers, it is challenging to provide full-body animation, and thus it is common to visualize only a part of the body (e.g., head and hands or upper body). The type of avatar visualization preferred by the user may depend on the type of task (Pastel et al. 2020) and also on the quality of the animations provided by the tracking system (Yun et al. 2023). In our work we developed a method to enhance the animation of full-body avatars to be used in immersive VR regardless of task requirements.

## 3 Avatar fitting framework

Our framework provides a pipeline to create fitted avatars to the specific dimensions of the user. To achieve this, the user needs three HTC trackers attached to the feet and back, an HMD, and a controller in each hand (see Fig. 2 left). Note that we locate the feet trackers on the user's feet as opposed to the ankle, since this allows us to have full control of the feet orientation to simulate, for example, walking on tiptoes. The user can see the virtual representation of those devices in the virtual world with a reflection in a virtual mirror (see Fig. 2 right). Next, the user follows a short sequence of exercises to automatically extract the location and dimensions of their joints and limbs (this set of exercises is explained in Sects. 3.2.1 and 3.2.2 and it takes around 60 s for the user to complete.). Once the avatar has been adjusted to the user's dimensions, the final step consists of walking inside a virtual body that matches the user's dimensions, making a T-pose inside the avatar and pressing a trigger to compute the exact offset between trackers and joints (Ponton et al. 2022). Figure 3 shows the steps required to set up the avatar with the correct user skeleton dimensions.

For the rest of the paper we will use the following notation:

- $\mathbf{p}(T_i)$: position of the tracking device $T$ for body part $i$.
- $\mathbf{p}(J_i)$: position of the joint $J$ for body part $i$.
- $\mathbf{h}(T_i)$: height of the tracking device $T$ for body part $i$.

- $\mathbf{h}(J_i)$: height of the joint $J$ for body part $i$.

### 3.1 Step 1: Automatic tracker identification

The first step consists of identifying and assigning each tracker to a body part. To determine where each device is placed on the user's body, we ask the user to stand on a T-pose, facing a virtual mirror. From the position of the six trackers, we compute a plane using a least square error fitting approach. The HMD forward vector and the vertical axis of the world are used to create a coordinate system to identify left and right devices. Then, knowing the relative positions of the trackers in this coordinate system, we uniquely identify each tracker to a body joint (see Algorithm 1 for details). Finally, we store the height of those trackers needed for the skeleton adjustment in Step 3, as $\mathbf{h}(T_i)$, where $i \in \{\text{HMD}, root, Rfoot, Lfoot\}$.

---

**Algorithm 1** Automatic tracker identification

---

**Input:** $\mathbf{D} = (\mathbf{p}(T_{HMD}), \mathbf{p}(T_1), \mathbf{p}(T_2), \mathbf{p}(T_3), \mathbf{p}(T_4), \mathbf{p}(T_5))$

1: // Plane that minimizes the squared distance to points in $D$
2: $(\mathbf{n}, d) \leftarrow FitPlane(D)$ // Plane defined using the normal vector and the distance to the origin
3: // Enforce forward HMD and Plane point at the same direction
4: $\mathbf{f} \leftarrow$ forward vector HMD
5: **if** $\mathbf{f} \cdot \mathbf{n} < 0$ **then** $\mathbf{n} \leftarrow -\mathbf{n}$
6: // Project devices' positions onto the plane
7: $\mathbf{P} \leftarrow \{Proj_{(\mathbf{n},d)}(\mathbf{x}) \mid \mathbf{x} \in \mathbf{D}\}$
8: // Define $(u, v)$ coordinate system
9: $\mathbf{v} \leftarrow (0, 1, 0)$ // Assuming Y is up
10: $\mathbf{u} \leftarrow \mathbf{v} \times \mathbf{n}$
11: $\mathbf{o} \leftarrow (\mathbf{P}_0 \cdot \mathbf{u}, \ \mathbf{P}_0 \cdot \mathbf{v})$ // Origin is the projected $\mathbf{p}(T_{HMD})$
12: $\mathbf{U} \leftarrow \{(\mathbf{x} \cdot \mathbf{u}, \mathbf{x} \cdot \mathbf{v}) - \mathbf{o} \mid \mathbf{x} \in \mathbf{P}\}$
13: // Assign indices to trackers
14: $HMD \leftarrow 0$ // HMD is known
15: **for** $(i \leftarrow 1; \ i \leq 5; \ i \leftarrow i + 1)$ **do**
16:     $(u, v) \leftarrow \mathbf{U}_i$
17:     **if** $T_i$ is Controller **then**
18:         **if** $u < 0$ **then** $Lwrist \leftarrow i$ **else** $Rwrist \leftarrow i$
19:     **else if** $T_i$ is HTC Tracker **then**
20:         **if** $v > 0.8$ **then** // Higher than threshold (meters)
21:             $root \leftarrow i$
22:         **else if** $u < 0$ **then**
23:             $Lfoot \leftarrow i$
24:         **else if** $u > 0$ **then**
25:             $Rfoot \leftarrow i$
26:         **end if**
27:     **end if**
28: **end for**

---

**Fig. 4** Synchronous arms exercise. The exercise consists of moving the arms in circular patterns (blue arrows) while keeping the arms as straight as possible (Color figure online)

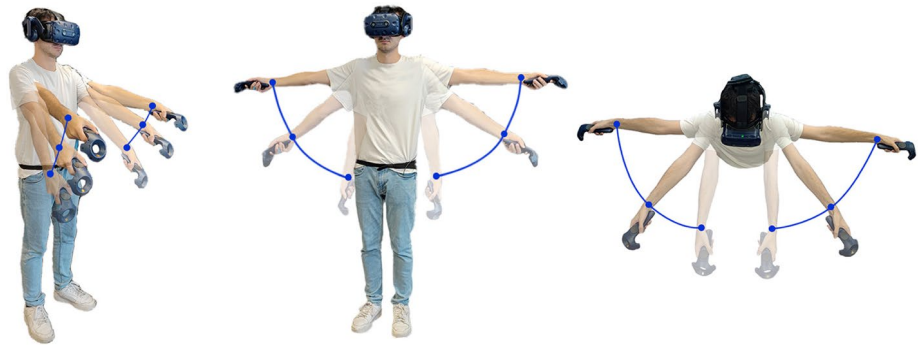

## 3.2 Step 2: Joint locations and limb dimensions

To fit an avatar to the user, it needs to be resized to match the user's dimensions, but more importantly, we need to correctly locate the relevant joint rotations. In our experience, most of the accuracy problems, that we observe when animating self-avatars with IK, are due to not having the joints' location correctly positioned with respect to the trackers. This can be frequently observed in the hand position, which appears to correctly hold the controller for certain arm positions but not for others, such as when the user moves the arms toward the sky. This happens because even though the size of the avatar's arm may be correct, the location of the avatar's shoulder joint is not. It is important to note that even if the avatar dimensions were correct, it will still not improve all possible movements, because the avatar's skeleton is a simplified version of a human one. So, for example, a person can move the shoulders a bit front or back, whereas a typical avatar skeleton does not have this degree of freedom as some bones are not represented (or are not animated by the IK solvers). Our method expects a humanoid avatar with at least three joints per limb, two in the spine and one for the neck.

In order to correctly map the user's body to the avatar, we need to infer the location of some of the user's body joints. Those locations can also be used to determine several body measures. The length of some parts of the body can be extracted directly from the distance between trackers, or from the height of the trackers when the participant is standing in T-pose. Since we want to keep the required number of trackers low, it is not possible to track every single joint individually. Therefore, starting with the basic configuration of HMD, two controllers and three VIVE trackers, we designed a set of exercises to infer the location of non-tracked joints. Since several body joints are centers of rotation when moving a limb, we can compute their location by calculating the center of a point cloud forming a sphere, which is captured by the tracker located at the end of the limb. We use a least squares solution for estimating the average center of rotation and the radius of the sphere (Gamage and Lasenby 2002). This method is used for the left and right arm and for the neck, as follows:

- Shoulders: position of the center of rotation $\mathbf{p}(C_{shoulder})$ of the point cloud obtained with the controllers tip positions $\mathbf{p}(T_{wrist})$ when moving it with the straight arms.
- Arm length: radius of the fitted sphere obtained from the hand controllers, which corresponds to the distance:

$$\mathbf{L_{arm}} = distance(\mathbf{p}(C_{shoulder}), \mathbf{p}(T_{wrist}))$$

- Neck: position of the center of rotation $\mathbf{p}(C_{neck})$ of the point cloud obtained with the HMD tracker positions $\mathbf{p}(T_{HMD})$ when performing head rotations.

In our experience, those body parts can be easily rotated while keeping the rest of the body still in a T-pose. Other parts of the body, such as the legs, are difficult to rotate while keeping the rest of the body still, and thus this method cannot be applied.

Other dimensions can be easily extracted from trackers, since they are accurately located on the user's body, or using the computed shoulders from the fitted sphere:

- Leg length—difference in height between the root and the feet trackers:

$$\mathbf{L_{leg}} = \mathbf{h}(T_{root}) - \frac{\mathbf{h}(T_{Lfoot}) + \mathbf{h}(T_{Rfoot})}{2}$$

- Torso length (up to shoulders)—difference in height between the computed shoulders and the root tracker:

$$\mathbf{L_{torso}} = \mathbf{h}(C_{shoulder}) - \mathbf{h}(T_{root})$$

- Shoulder to neck length—difference in height between the computed shoulder and the computed neck: the computed shoulders and the root tracker:

$$\mathbf{L_{neck}} = \mathbf{h}(C_{neck}) - \mathbf{h}(C_{shoulder})$$

- Neck to eyes length—difference in height between the HMD and the computed neck:

$$\mathbf{L_{eyes}} = \mathbf{h}(T_{HMD}) - \mathbf{h}(C_{neck})$$

**Fig. 5** Point cloud formed by the arms exercise for the fitting spheres algorithm that computes the shoulder joints

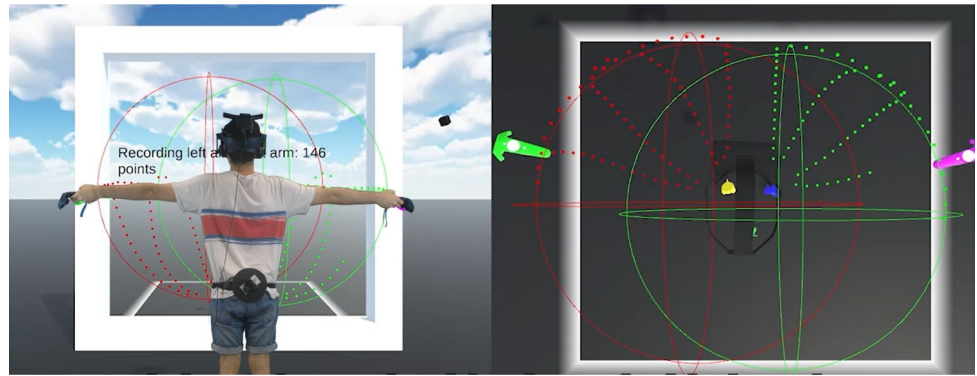

### 3.2.1 Shoulder joint extraction

Shoulders joint position $\mathbf{p}(C_{shoulders})$ is obtained by asking the user to do an exercise that consists of moving the arms in circles while keeping the arm as straight as possible. Both arms have to be moved synchronously to reduce the noise caused by other moving body parts (for example, a small movement of the spine will affect both shoulders' positions simultaneously by drifting slightly). Figure 4 shows the arm exercise movement with blue arrows. As the participant performs this exercise, we capture the point cloud of controller positions which will approximate the shape of a sphere (see Fig. 5). A least squares solution is used to estimate the center of rotation (Gamage and Lasenby 2002).

To ensure high precision when fitting the spheres, the algorithm has to satisfy two conditions to be completed (1) a minimum number of $n$ points, further than $d$ centimeters apart from each other, and (2) a convergence threshold $\tau$. See Algorithm 2 for details. After experimental training of the system, we set the following values $n = 70$, $d = 5$cm, and $\tau = 2$cm.

---

**Algorithm 2** Shoulder fitting

**Input: P:** captured fitting points
1: $n \leftarrow 70$, $d \leftarrow 0.05$, $\tau \leftarrow 0.02$ // distances in meters
2: **if** $\min_i \|\mathbf{P}_i - \mathbf{p}(T_{wrist})\| < d$ **then return** continue
3: $\mathbf{P}' \leftarrow \mathbf{P} \cup \{\mathbf{p}(T_{wrist})\}$
4: **if** $| \mathbf{P}' | < n$ **then return** continue
5: $\mathbf{c}' \leftarrow FittedSphere(\mathbf{P}')$
6: $\mathbf{c} \leftarrow FittedSphere(\mathbf{P})$
7: **if** $\|\mathbf{c}' - \mathbf{c}\| < \tau$ **then**
8: $\quad \mathbf{p}(C_{shoulder}) \leftarrow \mathbf{c}'$
9: $\quad$ **return** finish
10: **else**
11: $\quad$ **return** continue
12: **end if**

---

We considered using this method also for the wrists, but it was very hard for the participants to keep their arms completely still while rotating only the hands. Therefore, the wrist location is extracted directly from the controllers. To do so, the participant is asked to hold the controller with the tip touching their wrist (see Fig. 6 left).

### 3.2.2 Neck joint computation and head center estimation

Similarly, the neck position $\mathbf{p}(C_{neck})$ can be computed as the center of rotation of the head. Therefore, we perform a fitting sphere algorithm using the HMD position and ask the participant to perform an exercise that consists of moving the head up/down, left/right, and rotating in circles. Following the same algorithm as for the shoulders, we empirically found the following parameter values: $n = 60$, $d = 1.5$cm, and $\tau = 2$cm.

Since Unity's built-in IK does not provide an end-effector for the head, the self-avatar cannot perform movements such as bending forward or to the sides to reach for objects. Unfortunately, this affects the overall movements of the avatar, thus limiting the benefit of our fitted avatars. We thus included a forward kinematic method to compute the angle of rotation of the spine given by the root position $\mathbf{p}(T_{root})$ and the head's center position $\mathbf{p}(C_{head})$ (Ponton et al. 2022). The head position is a fixed point with respect to the HMD local coordinate system that has the least displacement when the user rotates the head wearing the HMD (Fig. 6 shows the movement of the head during the exercise with blue arrows). This point is computed during the neck exercise using the following method:

1. Define a 2D grid of cells located over the plane defined by the forward $\hat{\mathbf{f}}$ and right $\hat{\mathbf{r}}$ unit vectors of the HMD local coordinate system: $\mathbf{G}$ of resolution $n \times m$ with cells of size $s$.

2. Initialize the position of all cells on this grid in global coordinates as $\mathbf{G}_{ij}^0$

3. While the head is moving ($t \in [1, t_{end}]$) to perform the fitting sphere algorithm, compute for each cell of the grid the displacement between the current global position and its initial position, and keep track of the maximum displacement for each grid cell during the iterative head rotations.

$$\mathbf{M}_{ij}^t = \max(\mathbf{G}_{ij}^t - \mathbf{G}_{ij}^0, \ \mathbf{M}_{ij}^{t-1}),$$
$$t \in [1, t_{end}] \ \wedge \ \forall i, j \in \{n \times m\}$$

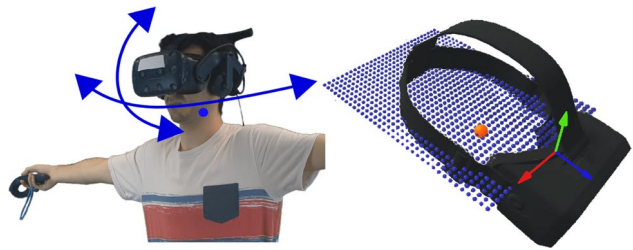

**Fig. 6** On the left, the center of rotation of the head (blue dot) is computed from the head exercise to determine the neck joint position. This exercise consists of moving the head upwards and sideways as shown with the blue arrows. On the right, the orange dot shows the point of minimum displacement over the plane which traverses the HMD when the head rotates. This point is used to bend the spine with forward kinematics (Color figure online)

4. When the head exercise finishes, we search for the cell with the smallest maximum displacement and keep its position as the center of the head. This point suffers the minimum displacement when the user rotates the head without bending the back.

$$(i^*, j^*) = \arg \min(\mathbf{M}_{ij}^{t_{end}}), \ \forall i, j \in \{n \times m\}$$

$$\mathbf{p}(C_{head}) = \mathbf{p}(T_{\mathrm{HMD}}) + s(-\hat{\mathbf{f}})i^* + s\hat{\mathbf{r}}\left(j^* - \frac{m}{2}\right)$$

### 3.3 Step 3: Skeleton adjustment

Once we know the distances between the required body joints, and their heights with respect to the floor, we can adjust the avatar's limbs' dimensions. In order to have a virtual avatar matching the user's body as close as possible, we follow a two-pass process: (1) uniform avatar scaling, and (2) skeleton tuning.

#### 3.3.1 Uniform avatar scaling

A uniform scaling is applied based on the user's height to preserve the avatar's proportions. Uniformly scaling the avatar ensures that we do not excessively deform its mesh in the skeleton tuning step. Since the participant's height is unknown, the scaling is done by computing the ratio between the participant's eyes height $\mathbf{h}(T_{\mathrm{HMD}})$ and the avatar's eye height $\mathbf{h}(J_{eye})$. Note that the uniform scaling step only guarantees a good match between the avatar and the user height. Any other limb dimension will depend on how close the user proportions were to the chosen avatar.

#### 3.3.2 Skeleton tuning

To resize the avatar bones based on the user's limb measurements, we focus on five bone chains: the two arm chains, the two leg chains and the spine chain. Our method stretches or shrinks the length of each of those chains while keeping the ratios between the bones' length for each chain. Please refer to Fig. 7 for the naming convention of the bones.

**Leg chain:** The avatar's lower and upper leg bones are resized according to the participant's leg height $\mathbf{L_{leg}}$.

**Arm chain:** Similarly, the lower and upper arm bones are resized to match the participant's arm length $\mathbf{L_{arm}}$.

**Spine chain:** First we adjust the shoulder width and then the spine length. The avatar's shoulder width is adjusted by scaling along the axis defined by the vector connecting both shoulder joints:

$$\mathbf{v} = \frac{\mathbf{p}(J_{Rshoulder}) - \mathbf{p}(J_{Lshoulder})}{\|\mathbf{p}(J_{Rshoulder}) - \mathbf{p}(J_{Lshoulder})\|}$$

According to this axis, shoulder joints are positioned so that they are at a distance $d$ apart. The distance $d$ is given by the distance between the shoulders' centers of rotation from Step 2:

$$d = distance(\mathbf{p}(C_{Rshoulder}), \ \mathbf{p}(C_{Lshoulder}))$$

Then we need to resize the vertical dimension of the bone chain comprising the avatar's spine, chest and upper chest bones, to match the avatar's shoulder height with the height of the shoulder centers of rotation $\mathbf{h}(C_{Lshoulder})$ and $\mathbf{h}(C_{Rshoulder})$.

The final step is to adjust the avatar's neck height to match the user's neck height. We first used the computed $\mathbf{p}(C_{neck})$

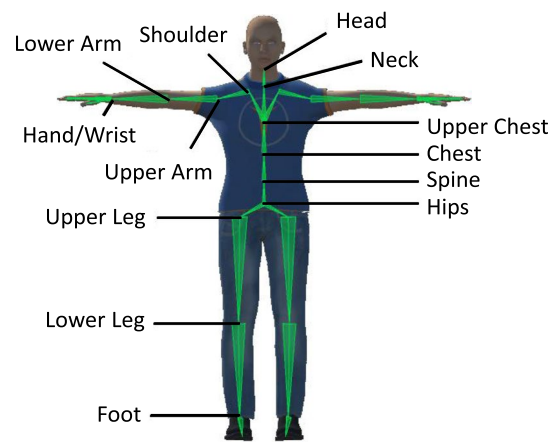

**Fig. 7** Avatar joints naming convention

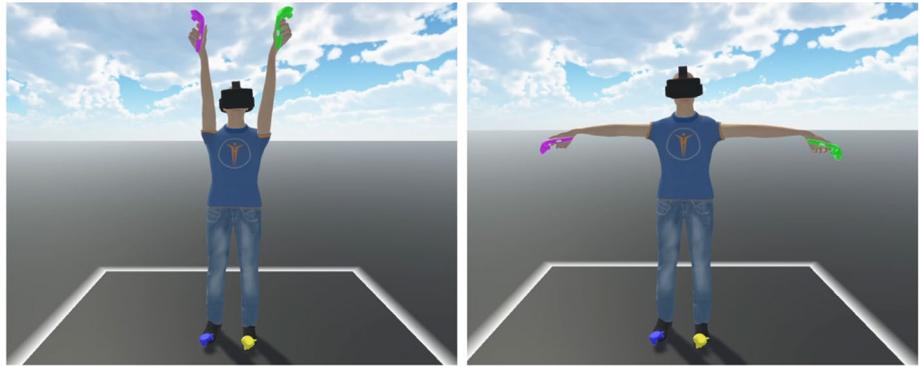

as the neck of the avatar. Unfortunately given the oversimplification of the neck bones in the avatar, there is not an exact correspondence for the joint $\mathbf{p}(J_{neck})$. Assigning the avatar's neck joint to the participant's computed neck would lead to an avatar with an extremely long neck from a perceptual point of view, whereas using the avatar's head bone would cause the opposite visual effect (a very short neck). So using our current avatar, the real participant's neck would be in between those two avatar joints, and thus, to keep the perceptual plausibility of the model, we resize those two bones to adjust the avatar's eye level with the participant's eye height. This solution offers a good compromise between the height of the avatar and the participant, while keeping a pleasant resizing of the avatar's head and neck. If the avatar model had more bones defined at the neck, it would be better to match the neck at its exact center of rotation. The center of the head computed in Sect. 3.2.2 is not needed for the skeleton adjustment, but for the forward kinematics needed to bend the spine.

### 3.4 Step 4: Walk-in-avatar for interactive coupling

For the final step, a virtual avatar with the correct user's dimensions is rendered in T-pose in front of a virtual mirror. The user can also see the virtual trackers attached to their real body. With this visual aid, the user is asked to walk into the avatar and stay in T-pose making sure that the virtual trackers appear located over the avatar, and press a trigger when there is a correct coupling between their body and the avatar.

This step computes the exact vector offset between the position of the lower back tracker and the avatar's root as well as their initial rotations. The same computation is performed for the feet trackers and the avatar's ankles. Those exact offsets will be used for the avatar animation, and are key to having a correct pose for the avatar at all times, as shown in Fig. 8 (details of the computation can be found in Ponton et al. 2022).

This step cannot be further automatized, because, without a correct overlaying of the user and the avatar, it would not be possible to compute the exact offset between the avatar's hip and the root tracker located on the participant's lower back. Most methods assume that the avatar hip corresponds to the root trackers, which leads to severe problems with the avatar's legs dimensions and results in avatars that may appear to walk on tiptoes or to have their knees always bent (Fig. 1).

Therefore, the user is responsible for positioning themselves so that the back tracker appears to be correctly aligned with the avatar's back, and similarly, feet trackers

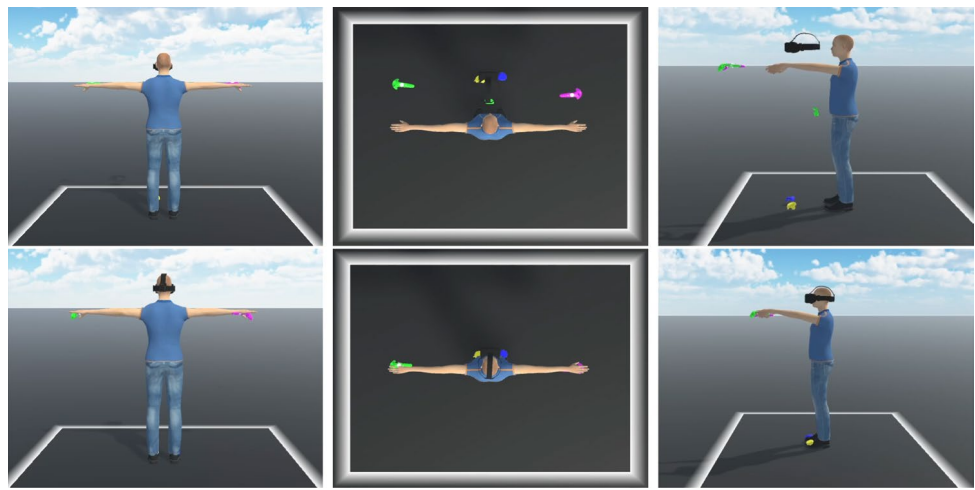

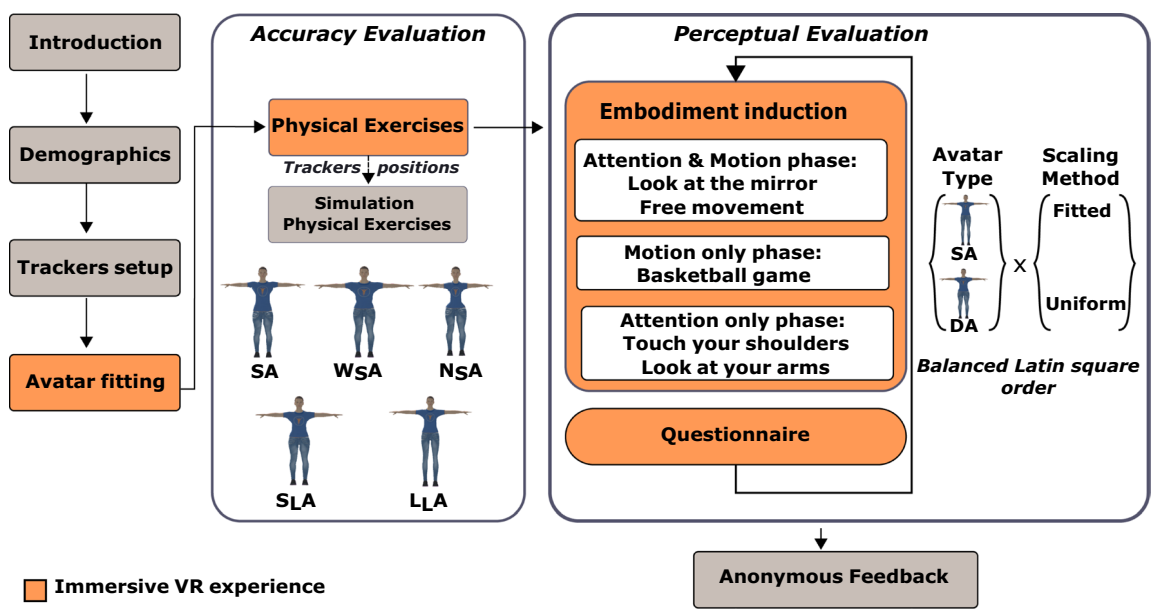

**Fig. 10** User study organization. The user study consists of two evaluations: *Accuracy Evaluation*, which consists of a quantitative statistical evaluation of a set of metrics; and a *Perceptual Evaluation*, which consists of a user perceptual evaluation of the *SoE*. The orange blocks are part of the immersive VR experience (Color figure online)

are correctly located over the avatar's feet. To help the user with this self-positioning, the user can use the mirror as a visual aid to get a front view, and a screen that shows the top view of the avatar simultaneously. The user can see the virtual avatar simultaneously with the virtual trackers. Any misalignment in this step (for instance the user locating himself/herself 10 cm in front of the virtual body) will obviously affect the offsets between the user body and its self-avatar (for example, when extending the arms forward, they would appear shorter and give the impression of not reaching the virtual controllers) see Fig. 9 top for a visual example of such situation. The result of a correct alignment within the fitted avatar is shown in Fig. 9 bottom.

# 4 Experimental evaluation

The goal of the evaluation is twofold: firstly we want to quantify the error in accuracy that appears when animating an avatar that has been obtained with uniform scaling as opposed to the fitted avatar provided by our framework (*Accuracy evaluation*); and secondly, we want to evaluate whether this error is perceptually noticeable for the user and whether or not it affects the sense of embodiment (*SoE*) (*Perceptual evaluation*). Figure 10 shows the pipeline for the experimental evaluation.

## 4.1 Data preparation: avatar types

To take into account the variability of human body dimensions, we build a database with several body proportions. We created five male and five female avatars with different body proportions (i.e., skeleton configurations) as shown in Fig. 11. All avatars were created with MakeHuman (which allows fine-tuning of body dimensions) and rigged with Mixamo.[4]

The five avatars (for each gender) created for the *Accuracy evaluation* are:

- **SA**: Standard Avatar with a shoulder width and ratio between lower and upper body following the *golden ratio* defined by Da Vinci in *the Vitruvian Man* (Magazù et al. 2019; Sheppard 2013).
- **$N_SA$**: Narrow Shoulder Avatar, obtained by reducing the shoulder width of **SA**.
- **$W_SA$**: Wide Shoulder Avatar, obtained by increasing the shoulder width of **SA**.
- **$L_LA$**: Long Leg Avatar, obtained by increasing the leg length of **SA**, while keeping the same height (thus reducing the upper body length).
- **$S_LA$**: Short Leg Avatar, obtained by decreasing the leg length of **SA**, while keeping the same height (thus increasing the upper body length).

---

[4] Adobe Mixamo: https://www.mixamo.com.

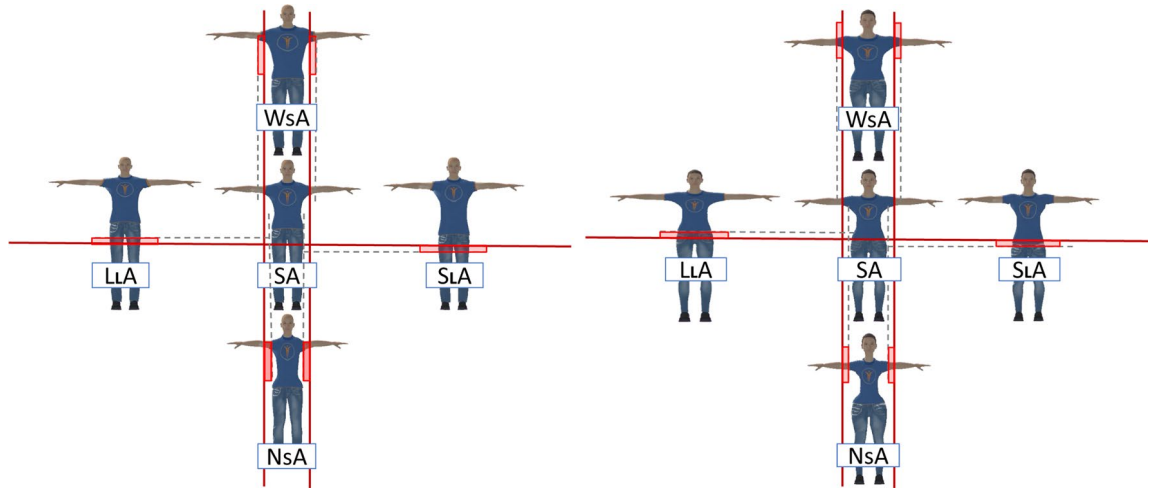

**Fig. 11** The avatars DB used in the *Accuracy Evaluation* experiment. Using the *MakeHuman* framework we built four different avatar proportions for male and female from a standard body proportions avatar (**SA**). The red squares show the difference in width/length between different body parts of the avatars (Color figure online)

Each avatar type will then be tested with the two scaling methods, (*Uniform* vs. *Fitted*) thus leading to 10 conditions. The *Uniform* method consists of exclusively applying a uniform scaling as described in Sect. 3.3.1, whereas the *Fitted* method consists of applying the complete framework as described in Sect. 3.

To reduce the number of avatars for the *Perceptual Evaluation*, we used only two avatar configurations: the Standard Avatar, **SA**, and a Different Avatar, **DA**. We generate **DA** for each participant by choosing the combination of shoulder width and upper/lower body ratio that provides the avatar with the proportions being as different as possible from the participant. The combination of those two avatar types (**SA** vs. **DA**) with the two scaling methods (*Uniform* vs. *Fitted*), led to 4 conditions (see Fig. 10).

### 4.2 Experimental design

Participants are first briefed about the steps of the experiment and the total duration. After signing a consent form and filling out a demographics questionnaire, they put on the three VIVE trackers, the HTC Vive PRO wireless HMD and take one controller in each hand. During the *Avatar fitting* step, the participant performs the exercises as shown in a video that is displayed within the immersive experience. During this part of the experiment, our pipeline computes the participant's skeleton dimensions. The participant then performs an accuracy evaluation followed by a perceptual evaluation.

#### 4.2.1 Accuracy evaluation

During this phase, the user performs a sequence of exercises while the pipeline extracts data to perform quantitative measurements for the five avatar conditions (**SA**, $L_LA$, $S_LA$, $W_SA$, $N_SA$), as shown in Fig. 11, and the two scaling methods *Uniform* and *Fitted*, thus testing 10 conditions. The avatar is not rendered while performing these exercises. The first exercise (*ArmsMovExercise*) consists of doing circles in the air with the arms completely stretched while standing in T-pose (similar to the first exercise of the *Avatar Fitting* setup described in Sect. 3.2.1). The second exercise (*LegsMovExercise*), consists of positioning the arms in an akimbo position while doing squats.

When performing these exercises, the application records all the positions and orientations of the trackers and controllers. Then we use this data to simulate the animation of each of the 10 avatar conditions while computing the offsets between each tracker and its corresponding joint. This allows us to use the exact same movements performed by the user for all five avatars. Section 5.1 explains the results obtained through a statistical analysis of the recorded data.

#### 4.2.2 Perceptual evaluation

For the *Perceptual evaluation*, participants are embodied in four avatar conditions (**SA_Fitted**, **SA_Uniform**, **DA_Fitted**, **DA_Uniform**) to evaluate the *SoE* based on avatar type and scaling method. For each condition, users carry out a few exercises and then fill out an embodiment questionnaire. Conditions were shown to participants following a balanced Latin square order to avoid learning/fatigue influencing the results. After the experiment, users can share their personal experiences by writing in a document.

To induce embodiment we ask participants to perform a sequence of steps, the first one involves moving freely in front of a mirror to observe how the virtual avatar mimics their movements (*Attention & Motion phase*) (Roth and Latoschik [2020]). For the second step, we ask the participants to play a simple basketball game (*Motion only phase*), which consists of grasping a colored ball using the right/left controller depending on the color of the ball, and dropping it inside a basket that appears randomly located within the virtual room. The purpose of this game is to make sure that participants pay attention to the avatar's arms. After approximately five minutes, the final step (*Attention only phase*) consists of asking the users to touch different parts of their body following our instructions. Finally, users answer an embodiment questionnaire, while still being immersed in the virtual experience.

The following questionnaire was answered by participants after experiencing each condition (using a Likert scale from 1 to 7), which is based on the proposed questionnaire by Roth and Latoschik ([2020]). It includes three main aspects of *Embodiment* (*change*, *ownership* and *agency*):

*CH1* "I felt like the form or appearance of my own body had changed."

*CH2* "I felt like the size (height) of my own body had changed."

*CH3* "I felt like the width of my own body had changed."

*CH4* "I felt like the size of my own arms had changed."

*CH5* "I felt like the size of my own legs had changed."

*OW1* "It felt like the virtual body was my body."

*OW2* "It felt like the virtual body parts were my body parts."

*OW3* "The virtual body felt like a human body."

*AG1* "The movements of the virtual body felt like they were my movements."

*AG2* "I felt like I was controlling the movements of the virtual body."

*AG3* "I felt like I was causing the movements of the virtual body."

Change is the average of questions $\{CH1, CH2, CH3, CH4, CH5\}$, *Ownership* is the average of questions $\{OW1, OW2, OW3\}$, and *Agency* is the average of questions $\{AG1, AG2, AG3\}$.

Some authors have included *Change* as a way of evaluating whether participants accept a different avatar as their own for studying the Proteus effect (Roth and Latoschik [2020]), or for experiments where they want to put participants in another person's shoes (e.g., to study racial or gender issues). This is not the case in our work, because all avatars look similar but with small differences in body proportions. Our goal is not to test if participants accept the body change, but whether participants notice differences in size between their real body and the avatar. Therefore, we define:

$$ResizeAccuracy = 7 - Change$$

$$SoE = (ResizeAccuracy + Ownership + Agency)/3$$

Small values of change prove a good avatar adjustment that is perceived as positive by the participant. Participants will score high when they consider that the virtual avatar has different proportions than them, thus implying a negative embodiment and the other way around.

### 4.3 Apparatus and participants

The software has been developed in Unity, and an HTC Vive PRO HMD wireless, two controllers and three trackers are used. A total of 27 subjects participated in the experiments, 17 males and 10 females, ranging between 18 and 62 years old. Participants were selected from our academic institution, encompassing faculty members, staff personnel, and master's/Ph.D. students, as well as individuals enrolled in higher education nondegree certificate (HNC) programs. 42% of the participants were familiar with VR applications. All participants exhibited unrestricted mobility without any observable movement impairments. For each embodiment question, subjects were asked to give a score on a Likert scale with 1 meaning very low and 7 very high. Our study used a within-subject design, so participants experienced all the conditions which followed a randomized design based on balanced Latin squares. Statistical analysis was performed with R. This study complies with UPC ethics requirements and regulations.

## 5 Results

We analyzed all the data collected in the experiments to assess how the scaling method used (*Uniform* vs. our *Fitted* method) and the different avatar types, influenced the accuracy of the avatar joints positioning with respect to the trackers (Sect. [5.1]) and the *Sense of Embodiment* (SoE) (Sect. [5.2]).

## 5.1 Accuracy evaluation

To evaluate quantitatively the quality of the avatar in terms of how well the virtual avatar was adjusted to the real body, we defined some anchor points in the avatar skeleton and measured their distance to the controllers while the participant performed the two exercises described in Sect. 4.2.1.

During the first exercise (*ArmsMovExercise*), we measured the distance between each hand controller and the corresponding wrist joint of the avatar. If the avatar's shoulder and arm length are correct, then, this distance should be small and stable while performing the exercises, otherwise, this distance grows when the controller is located in positions that are not reachable due to the inaccurate avatar size.

In the second exercise (*LegsMovExercise*), consisting in doing squats, the user places the controllers approximately at the top of each leg. During this exercise, we took two measures: the distance between each controller and the corresponding hip joint of the avatar, and also the knee angle. Therefore, the measurements taken in both exercises were:

- *Hand_{Dist}*: Distance between each controller and the corresponding wrist joint.
- *Leg_{Dist}*: Distance between each controller and the corresponding hip joint while being in the akimbo position.
- *Knee_{Ang}*: Angle of each knee joint.

Our hypotheses for the accuracy evaluation were:

**H1** Avatar hand position would be more accurate using the *Fitted* scaling method.

**H2** Avatar legs pose would be better adjusted using the *Fitted* scaling method.

A two-way repeated measures ANOVA (mixed model with the participant as a random effect) was performed using as response each distance variable (*Hand_{Dist}* and *Leg_{Dist}*). The considered fixed factors were the scaling method (*Uniform* vs. *Fitted*) and avatar type (5 levels: **SA, N_SA, W_SA, L_LA, S_LA**). Since different avatars are used for each gender, we performed the statistical analysis separately. Due to the presence of some outliers, the normality assumption was not thoroughly fulfilled, and thus, apart from the parametric approach we also applied the nonparametric aligned rank transform (ART) ANOVA approach (Wobbrock et al. 2011). We remark that both, the parametric and nonparametric approach, yielded the same results in terms of statistical significance; see "Appendix" for a detailed description of the complete statistical analysis including the nonparametric results in Tables 2 and 3.

Herein, we will mainly illustrate results with the classic repeated ANOVA analysis using the distances variable (*Hand_{Dist}* and *Leg_{Dist}*) .[5]

For each gender, the repeated ANOVA analysis of the *Hand_{Dist}*, yielded as significant the main factor method (*Uniform* vs *Fitted*), the avatar type (5 models) and also their interaction (all *p* values < 0.0001); see Table 2 (top) in "Appendix". Additionally, we computed Cohen's *f* effect sizes (Cohen 1988), which can be interpreted as *very small* ($f \geq 0.01$), *small* ($f \geq 0.2$), *medium* ($f \geq 0.5$), *large* ($f \geq 0.8$), *very large* ($f \geq 1.2$), or *huge* ($f \geq 2.0$) (Sawilowsky 2009). Regarding male participants, we observed a *huge* effect on the scaling method factor ($f = 2.30$), a *large* effect on the interaction between factors ($f = 0.84$), and a *small* effect for the avatar type factor ($f = 0.43$). When examining female participans, we observed a *large* effect in all cases: scaling method ($f = 1.11$), avatar type ($f = 0.92$), and interaction ($f = 0.84$). Figure 12 (left) also suggests that participants' bodies were better adjusted when using the *Fitted* method.

Post hoc comparisons were carried out, and the obtained effects (*Uniform* vs. *Fitted*) for all the avatar type configurations are displayed in Fig. 12 (right) and in Table 3 (top). There is only one configuration ($S_LA$-female) where the *Uniform* method outperforms the fitted method, but, as observed in the forest plot, the effect value is very close to 0.0. The effect size for *Hand_{Dist}* ranges from 4.0 cm. to 14.0 cm for males and from 2.6 to 8.0 cm for females. These results support our hypothesis **H1**.

Regarding the *Leg_{Dist}* variable, which is in charge of analyzing the low body when performing the *LegsMovExercise*, we also found as significant the main factor method (*Uniform* vs *Fitted*), the avatar type (5 models) and also their interaction (all p-values < 0.0001); see Table 2 (bottom). When examining the male participants, we found a *large* effect size for the interaction between factors ($f = 1.0$), and *medium* effects for the scaling method factor ($f = 0.77$) and avatar type factor ($f = 0.66$). When considering female participants, we observed a *large* effect for the scaling method factor ($f = 1.07$), and *medium* effects for both the avatar type factor ($f = 0.55$) and the interaction between the two factors ($f = 0.72$). Corresponding post hoc comparison effects are displayed in Fig. 15 (right) and in Table 3 (bottom). These results support our hypothesis **H2**.

Additionally, we analyze the knee angle variable, and we observed some interesting behavior. Figure 13 shows the time plot of the knee angle when a participant is performing the exercise for the different avatar types under both scaling methods. If we observe the variability in the knee angle value, it is larger when using the *Uniform* method than when using the *Fitted* method. This is due to the fact that when using the *Uniform* scaling, there is a larger variability in leg length, and thus in the correspondence between the user and the avatar knee bending. In contrast, for the *Fitted* method, where the avatars' leg length is the same as the user, then for all avatars we observe a similar knee angle during the squatting exercise. This result also supports **H2**.

---

[5] Distance variables (*Hand_{Dist}* and *Leg_{Dist}*) stand for left body side. Results corresponding to the right side were similar in terms of statistical significance and thus omitted here.

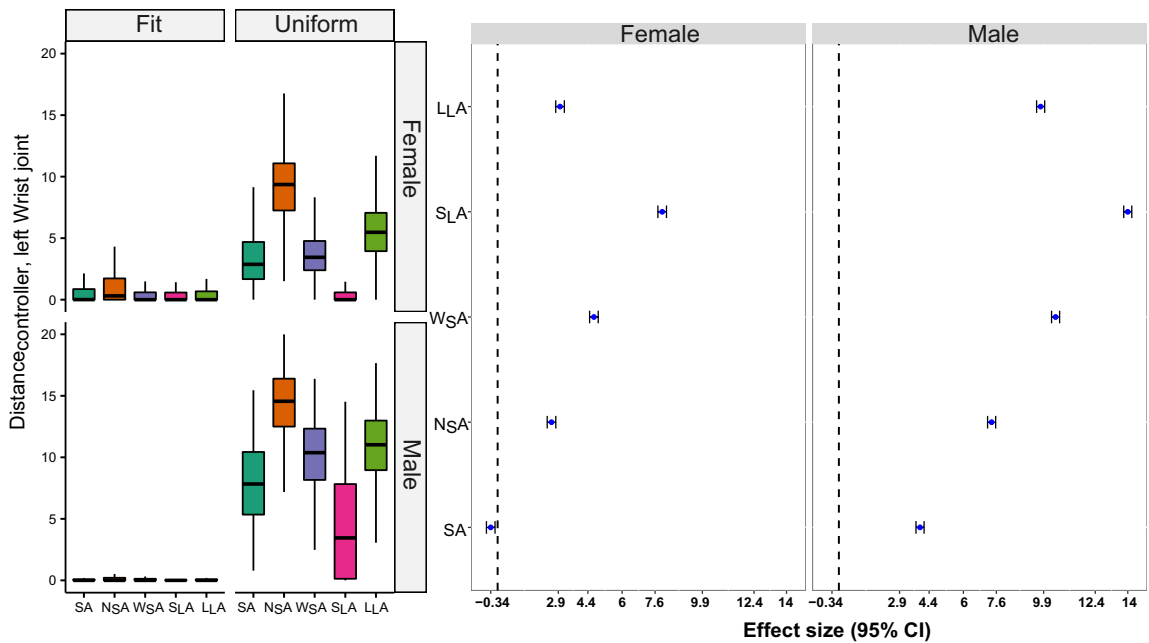

**Fig. 12** Results summary for the *ArmsMovExercise*: (left) Boxplot of the distance between the controller and the left wrist joint grouped by gender and method. (right) Forest plot (for each gender) of post hoc comparison effects and corresponding 95% CIs for the five avatar type configurations

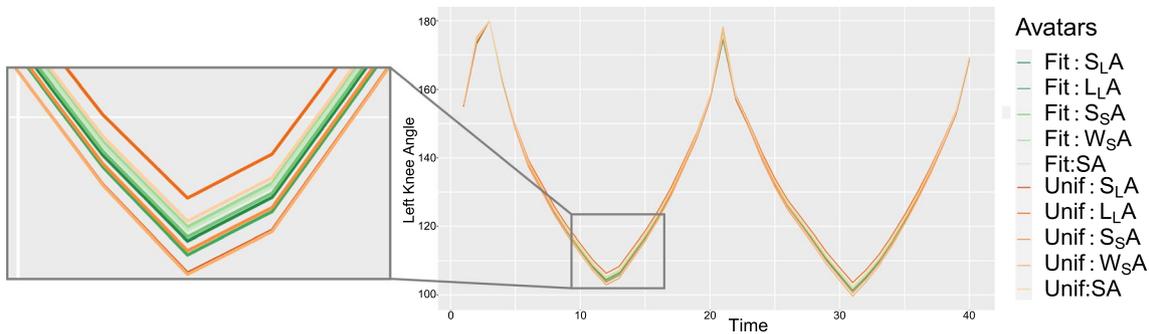

**Fig. 13** Illustration of results for the *LegsMovExercise*: Time plot of the knee angle for the different avatar types in both methods for a single participant. As shown the variability in the angle value is larger in the *Uniform* scaling method than in the *Fitted* scaling method

## 5.2 Perceptual evaluation

We carried out the perceptual experiment described in Sect. 4.2.2 to study the impact of avatar type and scaling method on the *SoE*. Answers from the questionnaires were structured into three groups (*ResizeAccuracy*, *Ownership*, and *Agency*). Our main hypothesis was:

**H3** Embodiment ratings would be higher when using a *Fitted* avatar.

A two-way ANOVA was performed to analyze the effect of avatar type (**SA** vs. **DA**) and scaling method (*Fitted* vs.

*Uniform*) on *SoE*. Results were not statistically significant and not shown.

Since, as planned in our experiment, **DA** had been chosen as the most different avatar in terms of body proportions to each participant, we then restricted our statistical analysis to this condition. The corresponding paired t-test was performed comparing the scaling method for the SoE (and its components). The obtained results were statistically significant (higher) only for SoE and the *Ownership* component; (SoE: $t = 2.34, df = 21, p < 0.029$; Ownership: $t = 2.72, df = 21, p < 0.013$). The normality assumption for the differences was verified by the Shapiro–Wilk normality test (SoE: $W = 0.94, p < 0.17$; Ownership: $W = 0.95, p < 0.36$). Figure 14 includes the summary of this analysis (the **DA$_{Uniform}$** vs. the **DA$_{Fitted}$**). These results support our hypothesis **H3**.

To further justify the necessity of our *Fitted* scaling method, when restricting the statistical analysis to the *Uniform* scaling method, a paired t-test was performed comparing the *ResizeAccuracy* in both conditions: **SA** versus **DA**. The obtained results showed significant and higher values for **SA** ($t = 2.54, df = 21, p < 0.02$). This indicates that when participants use an avatar with very different proportions to them, they will notice the mismatch in dimensions if a uniform scaling is applied. The normality condition of the differences was tested using the Shapiro–Wilk normality test: $W = 0.94, p < 0.208$.

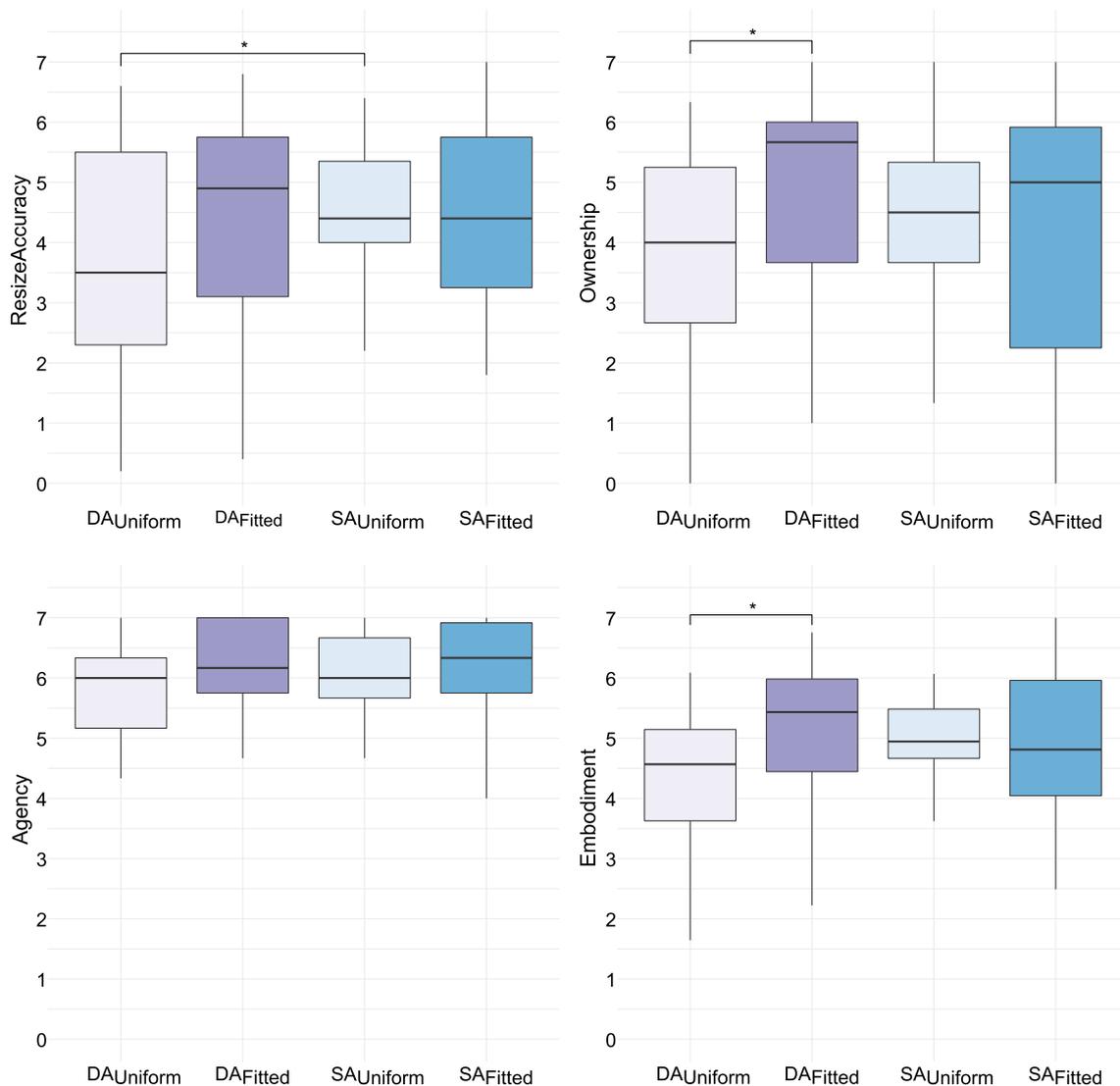

**Fig. 14** Boxplot of questionnaire ratings for *ResizeAccuracy*, *Ownership*, *Agency*, and *Embodiment* (SoE) for the different configuration analyzed: the **SA** and **DA** with respect to the scaling method used (*Uniform* vs. *Fitted*)

**Table 1** Descriptive summary of the distance–response variables

| Method | Variable | Avatar type | | | | |
|---|---|---|---|---|---|---|
| | | $SA$ | $S_{LA}$ | $L_{LA}$ | $N_{SA}$ | $W_{SA}$ |
| *Female* | | | | | | |
| Uniform | $Hand_{Dist}$ | 2.97 ± 2.43 | 0.37 ± 0.87 | 5.0 ± 2.33 | 8.62 ± 2.86 | 3.33 ± 2.11 |
| Fitted | | 0.96 ± 1.65 | 0.91 ± 1.66 | 0.94 ± 1.65 | 1.29 ± 1.84 | 0.92 ± 1.61 |
| Uniform | $Leg_{Dist}$ | 11.51 ± 2.42 | 12.57 ± 2.70 | 11.05 ± 1.97 | 11.89 ± 2.56 | 15.96 ± 3.14 |
| Fitted | | 8.30 ± 2.35 | 8.17 ± 2.48 | 11.56 ± 2.54 | 8.12 ± 2.48 | 8.90 ± 2.38 |
| *Male* | | | | | | |
| Uniform | $Hand_{Dist}$ | 7.54 ± 3.20 | 4.17 ± 3.89 | 10.61 ± 2.7 | 14.23 ± 2.47 | 9.96 ± 2.8 |
| Fitted | | 0.24 ± 0.66 | 0.21 ± 0.64 | 0.22 ± 0.61 | 0.26 ± 0.68 | 0.23 ± 0.65 |
| Uniform | $Leg_{Dist}$ | 9.4 ± 2.0 | 12.69 ± 2.57 | 10.07 ± 1.68 | 11.86 ± 2.44 | 13.50 ± 2.50 |
| Fitted | | 9.18 ± 1.78 | 9.20 ± 1.98 | 11.98 ± 2.39 | 9.47 ± 2.04 | 9.55 ± 1.85 |

For each gender, the mean and standard deviation of the variables $Hand_{Dist}$ and $Leg_{Dist}$ for the five avatar types in the two scaling methods are shown

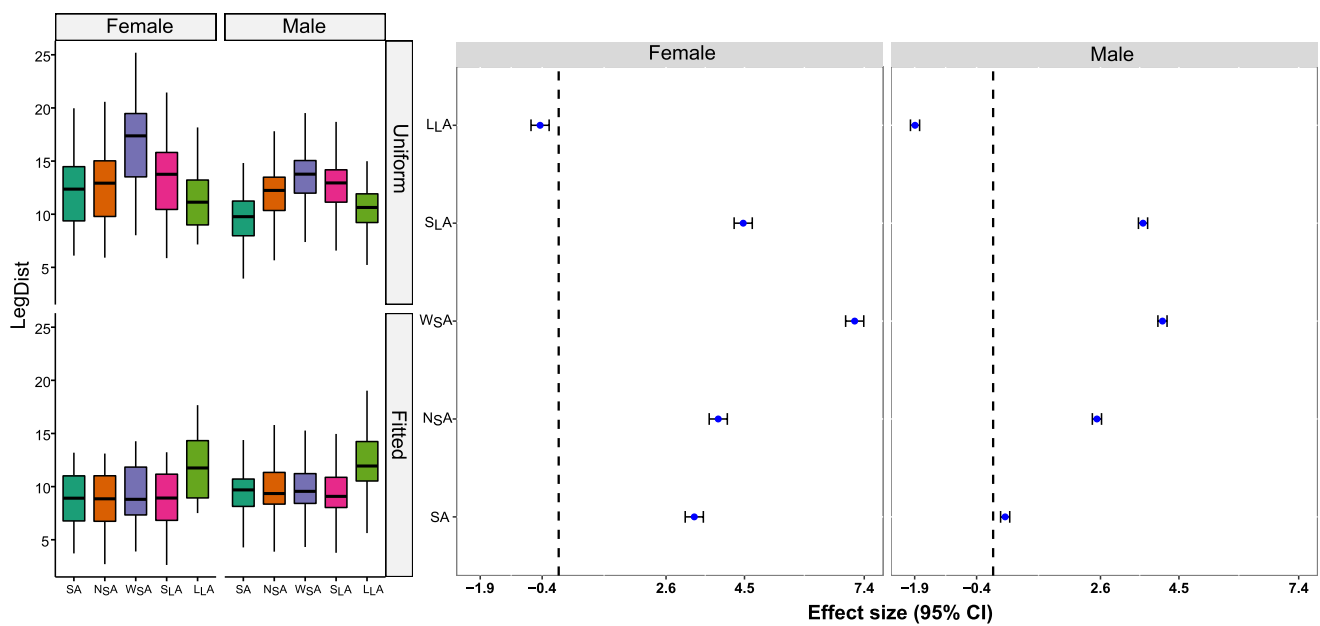

**Fig. 15** Results summary for the *LegsMovExercise*. *Top left:* Boxplot of the distance between the controller and the hip joint grouped by gender and method. *Top right:* forest plot (for each gender) of post hoc comparison effects and corresponding 95% CIs for the five avatar type configurations

## 6 Discussion

The presented framework provides an automatic fitting of avatars to better match the user dimensions. We have tested both quantitatively and qualitatively our results to evaluate the quality of our method when compared against a simple uniform scaling based on user height.

Regarding the positioning of the avatar hand with respect to the controllers, our results show that using our *Fitted*

scaling method the accuracy is improved regardless of the avatar type being used. Similarly, the accuracy of the distance between the avatar hip joint and the controllers in the akimbo position was also improved when using our *Fitted* scaling method.

It is interesting to highlight that the participants' leg movement when looking at the knee angle, was well reproduced for all the avatar types regardless of the scaling methods. The reason for this is that our exact offset computation between the back tracker and the hip joint, allows the avatar pose to be correctly initialized regardless of the avatar type or scaling method (i.e., they are all initially in a T-pose). Then when the participant bends the knees, all avatars behave correctly by bending the knees in sync with the user, and when the participant straightens the leg, all avatars return to a straight leg position. The errors in leg size simply produce differences in the knee angle which can be observed in a quantitative evaluation, but are not perceivable by the user. Ideally, we would like to compare the knee angle against the real participant knee angle, but this was beyond the scope of this project since it would require specific hardware to measure it automatically (for example a full motion capture suit).

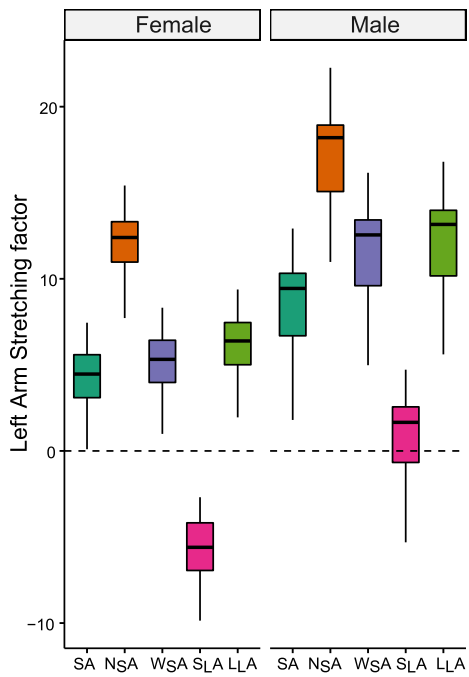

**Fig. 16** Boxplot of the left arm stretching scaling factor grouped by gender and avatar type

When we only apply uniform scaling, it is likely that the arm's length does not match the user dimensions (since the ratios of the initial avatar remain unchanged). Therefore, for those avatars with shorter arms, the user will observe that the avatar stays in T-pose for a few seconds after the user starts moving the arms toward the body. This happened to most of our avatars, since the stretching needed to adjust the length of the user's arm was positive (see Fig. 16). To measure the correctness of the elbow angle, we should also compare it against ground truth data from a full motion capture suit.

**Table 2** Comparison of classic repeated ANOVA and nonparametric ART ANOVA

| | Classical ANOVA | | | | ART ANOVA | | | |
|---|---|---|---|---|---|---|---|---|
| | numDF | denDF | F-value | p-value | numDF | denDF | F-value | p-value |
| *Hand$_{Dist}$ for female Gender* | | | | | | | | |
| Method | 1 | 7981 | 9785.024 | < .0001 | 1 | 7981 | 13,289.4 | < .0001 |
| Type | 4 | 7981 | 1691.555 | < .0001 | 4 | 7981 | 2482.4 | < .0001 |
| Method:Type | 4 | 7981 | 1408.277 | < .0001 | 4 | 7981 | 2141.0 | < .0001 |
| *Hand$_{Dist}$ for male Gender* | | | | | | | | |
| Method | 1 | 13,574 | 71804.31 | < .0001 | 1 | 13,574 | 59541.7 | < .0001 |
| Type | 4 | 13,574 | 2509.78 | < .0001 | 4 | 13,574 | 2844.0 | < .0001 |
| Method:Type | 4 | 13,574 | 2416.56 | < .0001 | 4 | 13,574 | 2772.6 | < .0001 |
| *Leg$_{Dist}$ for female Gender* | | | | | | | | |
| Method | 1 | 7981 | 9211.285 | < .0001 | 1 | 7981 | 9234.4 | < .0001 |
| Type | 4 | 7981 | 594.423 | < .0001 | 4 | 7981 | 578.67 | < .0001 |
| Method:Type | 4 | 7981 | 1028.879 | < .0001 | 4 | 7981 | 1031.71 | < .0001 |
| *Leg$_{Dist}$ for male Gender* | | | | | | | | |
| Method | 1 | 13,574 | 8057.512 | < .0001 | 1 | 13,574 | 7748.0 | < .0001 |
| Type | 4 | 13,574 | 1480.882 | < .0001 | 4 | 13,574 | 1300.1 | < .0001 |
| Method:Type | 4 | 13,574 | 3373.423 | < .0001 | 4 | 13574 | 3045.5 | < .0001 |

`Method` refers to the scaling method and `Type` refers to the avatar type

**Table 3** Post hoc comparisons for the distance–response variables ($Hand_{Dist}$ and $Leg_{Dist}$) of the *Accuracy Evaluation* experiment; all results were statistical significant

| | Female | | Male | |
| --- | --- | --- | --- | --- |
| | Effect size | 95%CI | Effect size | 95%CI |
| $Hand_{Dist}$ | | | | |
| $SA$:Uniform vs. Fitted | 2.613 | (2.40, 2.82) | 7.40 | (7.21, 7.60) |
| $S_{LA}$:Uniform vs. Fitted | − 0.33 | (− 0.54, − 0.13) | 3.93 | (3.74, 4.13) |
| $L_{LA}$:Uniform vs. Fitted | 4.67 | (4.46, 4.88) | 10.510 | (10.31, 10.70) |
| $N_{SA}$:Uniform vs. Fitted | 7.98 | (7.77, 8.19) | 14.010 | (13.81, 14.20) |
| $W_{SA}$:Uniform vs. Fitted | 3.02 | (2.82, 3.23) | 9.784 | (9.59, 9.98) |
| $Leg_{Dist}$ | | | | |
| $SA$:Uniform vs. Fitted | 3.28 | (3.06, 3.50) | 0.29 | (0.18, 0.40) |
| $S_{LA}$:Uniform vs. Fitted | 4.47 | (4.25, 4.69) | 3.63 | (3.52, 3.52) |
| $L_{LA}$:Uniform vs. Fitted | − 0.45 | (− 0.67, − 0.23) | − 1.89 | (− 2.00, − 1.78) |
| $N_{SA}$:Uniform vs. Fitted | 3.86 | (3.64, 4.08) | 2.51 | (2.40, 2.62) |
| $W_{SA}$:Uniform vs. Fitted | 7.17 | (6.95, 7.39) | 4.10 | (3.99, 4.21) |

These values are shown in the corresponding forest plots in Figs. 12 and 15

The perceptual study shows that when using the *Uniform* scaling method, users perceived the *Resize Accuracy* as higher for **SA** than for **DA**. This implies that uniform scaling will lead to perceivable size errors if the avatar used has different proportions from the user. Regarding ownership and *SoE* for the DA (avatar with proportions different to the user), participants rated significantly higher scores for the *Fitted* scaling than for *Uniform*. Therefore, our method to compute fitted avatars can reduce the perceivable errors that appear when using an avatar that is not proportionally similar to the user, and lead to higher Ownership and *SoE*. However, the qualitative differences perceived by the participants were smaller than the quantitative differences. The reasons for this could be twofold (1) the tasks that the users had to perform did not require careful positioning of the body limbs or accurate manipulation of virtual objects, and thus errors in the positioning of the end-effectors were not highly noticeable for the users, and (2) our current method correctly resizes each limb, but in some cases artifacts may appear with the mesh resizing or the rigging, thus leading to other noticeable visual artifacts that may impair the benefits of having accurate skeleton dimensions.

It is important to note that our method assumes that the user can follow correctly the rules and exercises indicated, since the quality of the results depends to a large extent on how well the user is following those exercises. During our experiments, all participants performed the exercises correctly and the automatic extraction was quickly achieved. However, people with limited mobility may not be able to perform the arm or head rotation correctly and thus the automatic extraction of joints could be less accurate.

## 7 Conclusion

We have presented a method to automatically compute accurate body dimensions and joint locations from a user, to then adjust the skeleton of a self-avatar to minimize the errors in poses and position of end-effectors during animation. By computing the most accurate location of joints and replicating their positions correctly in the avatar, our framework is able to improve the final body positions calculated by an inverse kinematics (IK) solver. Our method improves the final accuracy regardless of the avatar type being used, and thus facilitates the use of avatars for immersive VR applications using low-cost setups, such as an HTC VIVE system with three additional trackers.

## Appendix: complementary statistical analysis results

This appendix collects all statistical results of the *Accuracy Evaluation* experiment described in Sect. 5.1. Table 1 shows the descriptive analysis of the distance–response variables ($Hand_{Dist}$ and $Leg_{Dist}$).[6]

Figure 12 and Fig. 15 (top) show that the bodies of the participants were better adjusted when using the *Fitted* scaling method. A further justification of the differences observed can be explained when analyzing the stretching performed in each avatar. We have collected the avatar stretching (scaling) applied for all five avatars and for all participants; see Fig. 16. We observed that the avatars for male participants were always larger than the avatars for female participants, being $N_{SA}$ the avatar with the greatest scaling (arms were scaled 23 cm, as opposed to an average of 12 cm for the rest of the avatars) (Tables 2, 3).

---

[6] Distance variables ($Hand_{Dist}$ and $Leg_{Dist}$) stand for left body side. Results corresponding to the right body side were similar in terms of statistical significance and were thus omitted here.

**Supplementary Information** The online version contains supplementary material available at https://doi.org/10.1007/s10055-023-00821-z.

**Acknowledgements** This work was funded by the Spanish Ministry of Science and Innovation (PID2021-122136OB-C21). Jose Luis Ponton was also funded by the Spanish Ministry of Universities (FPU21/01927).

**Funding** Open Access funding provided thanks to the CRUE-CSIC agreement with Springer Nature.

**Data availability** Availability of data and materials data and related materials are available upon reasonable request to the corresponding author.

## Declarations

**Conflict of interest** The authors have no relevant financial or nonfinancial interests to disclose